\documentclass[11pt,twoside]{article}
\usepackage{graphicx,epsfig,natbib,epstopdf}
\usepackage{CS18}
\usepackage{url}
\usepackage{natbib,hyperref,amsmath}

\newcommand{\rearth}{\ensuremath{\mbox{R}_{\rm{Earth}}}}

\newcommand{\eg}{{\it e.g.}}

\newcommand{\kepler}{\it Kepler}
\def\rsun{R_{\odot}}
\def\teff{T_\mathrm{eff}}
\def\mk{\mathrm{M}_\mathrm{K}}
\def\feh{\mathrm{[Fe/H]}}

\begin{document}
%
%
%
\title{Touchstone Stars:\\ Highlights from the Cool Stars 18 Splinter Session}
%
%
\author{Andrew W. Mann$^{1,2}$, Adam Kraus$^{2}$, Tabetha Boyajian$^{3}$, Eric Gaidos$^{4}$, Kaspar von Braun$^5$, Gregory A. Feiden$^6$, Travis Metcalfe$^7$, Jonathan J. Swift$^8$, Jason L. Curtis$^9$, Niall R. Deacon$^5$, Joseph C. Filippazzo$^{10}$, Ed Gillen$^{11}$, Neda Hejazi$^{12}$, Elisabeth R. Newton$^{13}$}
\affil{$^1$Harlan J. Smith Fellow; amann@astro.as.utexas.edu}
\affil{$^2$Department of Astronomy, The University of Texas at Austin, Austin, TX 78712, USA}
\affil{$^3$Department of Astronomy, Yale University, New Haven, CT 06511, USA}
\affil{$^4$Department of Geology \& Geophysics, University of Hawai'i, 1680 East-West Road, Honolulu, HI 96822, USA}
\affil{$^5$Max-Planck-Institute for Astronomy (MPIA), K\"{o}nigstuhl 17, 69117 Heidelberg, Germany}
\affil{$^6$Department of Physics and Astronomy, Uppsala University, Box 516, SE-751 20 Uppsala, Sweden}
\affil{$^7$Space Science Institute, 4750 Walnut Street, Suite 205, Boulder, CO 80301, USA}
\affil{$^8$Department of Astrophysics, California Institute of Technology, MC 249-17, Pasadena, CA 91125, USA}
\affil{$^9$NSF Graduate Research Fellow; Department of Astronomy \& Astrophysics, The Pennsylvania State University, University Park, PA 16802, USA}
\affil{$^{10}$Department of Physics, City University of New York Graduate Center, 365 Fifth Avenue, New York, NY 10016, USA}
\affil{$^{11}$Sub-department of Astrophysics, Department of Physics, University of Oxford, Keble Road, Oxford OX1 3RH, UK}
\affil{$^{12}$Department of Physics and Astronomy, York University, North Yorkinterferometry, ON M3J 1P3}
\affil{$^{13}$Harvard-Smithsonian Center for Astrophysics, 60 Garden Street, Cambridge, MA 02138, USA}

\clearpage
\begin{abstract}
We present a summary of the splinter session on ``touchstone stars'' -- stars with directly measured parameters -- that was organized as part of the Cool Stars 18 conference. We discuss several methods to precisely determine cool star properties such as masses and radii from eclipsing binaries, and radii and effective temperatures from interferometry. We highlight recent results in identifying and measuring parameters for touchstone stars, and ongoing efforts to use touchstone stars to determine parameters for other stars. We conclude by comparing the results of touchstone stars with cool star models, noting some unusual patterns in the differences.
\end{abstract}
%
%
%
%
%
\section{Introduction}
Touchstones are stones traditionally used to verify the authenticity and purity of precious metals. By analogy, touchstone stars are those for which fundamental parameters - such as effective temperature, mass, or radius - have been directly and precisely measured. Touchstone stars may, for example, be eclipsing binaries, stars with asteroseismic measurements, or objects that have been targeted with interferometry. These stars can be used to test models of stellar evolution and atmospheres, and the properties of other stars can be determined by empirical comparison with these touchstone stars. Such stars have become increasingly useful and popular for studying the properties of cool stars, particularly late-type and young stars for which methods to measure stellar parameters are not well-established.

Direct measurements of fundamental parameters are not within reach for the vast majority of stars, and characterizing stars becomes even more challenging for stars without parallaxes. For surveys like {\it Kepler}, we know the distance to only a small percentage of the targets, but precise stellar parameters are required for accurate planet parameters. Models can provide some constraints, but are less well tested for evolved, very hot, very cold, and young stars because we cannot use the Sun as a reference. For these stars, the models have systematic errors that need to be determined for models to be useful for stellar characterization. 

Different kinds of touchstone stars provide different insight into stellar properties. By using stellar oscillations as seismic waves, asteroseismology probes stellar densities (subject to the accuracy of the effective temperature). Pairs or clusters of stars from a common birth cloud or molecular core provide sets of stars with similar (or identical) ages and metallicities. Interferometry can be used to measure the angular diameter of a star, which, combined with a measurement of the star's bolometric flux and distance, yields the stellar radius and effective temperature. Photometric and spectroscopic observations of eclipsing binaries offer precise stellar radii and masses. 

The growing use of cool touchstone stars has motivated this in-depth look at how their properties were determined, and how they can be used to study stellar properties more generally. Here we summarize some of the major contributions on cool touchstone stars presented in the Cool Stars 18 splinter session. Each section represents a contribution from one of the participants, covering the identification of new touchstones, establishing their properties, some of their uses, and conclude with a comparison of results from late-type touchstone stars to predictions from models. 

\section{Benchmark companions to nearby stars from Pan-STARRS 1}
Wide binaries provide key astrophysical laboratories for studying the coolest stars. While wide multiple systems as a population can act as a metric for the star formation process, it is their utility as benchmark systems that has been most valuable in recent years. Very cool/ultracool companions to well characterized main sequence stars can have their age and metallicity inferred from the primary. These are then be used along with the luminosity of the companion to calculate an effective temperature using evolutionary models. This can be compared to an effective temperature derived from atmospheric model fits to the companion's observed spectrum. A literature review in \cite{Deacon2012} found that most recent atmospheric models reproduce evolutionary model effective temperatures well but that some systems \citep[e.g., the Epsilon Indi system, ][]{King2010,Liu2010} remain outliers.

The ideal tool for identifying the coolest wide companions to nearby stars is a red-sensitive multi-epoch survey such as Pan-STARRS\,1 \citep{Kaiser2002}. This allows the identification of faint, red objects with common proper motion to their proposed primary. In conducting a search for wide companions in Pan-STARRS\,1 data, \citet{Deacon2014} first compiled a list of nearby candidate primaries from Hipparcos \cite{vanLeeuwen2007} and other proper motion catalogues \citep{Salim2003,Lepine2005,Lepine2011}. A region around these targets was examined for objects with common proper motion. After visually inspecting candidates the selection was refined using photometry from 2MASS \citep{Skrutskie2006}, UKIDSS \citep{Lawrence2007}, VISTA VHS \citep{Emerson2010} and UKIRT UHS (Dye et al., in prep.). After this objects were followed-up spectroscopically in the $JHK$ bands using SpeX on the IRTF \citep{Rayner2003}.

This survey yielded 61 new discoveries as well as the spectral classification of 27 previously known objects. In total this increased the sample of wide (projected separation$>$300\,AU) late M dwarfs by 88\% and the number of L dwarf companions in the same separation range by 96\% \cite{Deacon2014}. This is the largest sample of wide ultracool companions ever discovered. In addition the companion to HIP 6407 was resolved into a tight 0.13 arcsecond L1+T3 binary. While hierarchical triples such as this one are ideal benchmark objects, the orbital period of the two components of the secondary is over 300 years making direct mass determination impractical.

\section{Radii of Planet Hosts from Interferometry}
The combination of interferometry data with trigonometric parallax values and spectral energy distribution fitting based on literature photometry allows for direct determination of stellar radius, effective temperature, and luminosity. Observations through the CHARA Interferometric Array over the course of the past $\ge5$ years have significantly improved the properties of main-sequence stars and exoplanet hosts with a particular concentration on cool stars. The survey's primary research goals and applications are to use the stars as touchstones to achieve the following:

\begin{figure}
\plotone{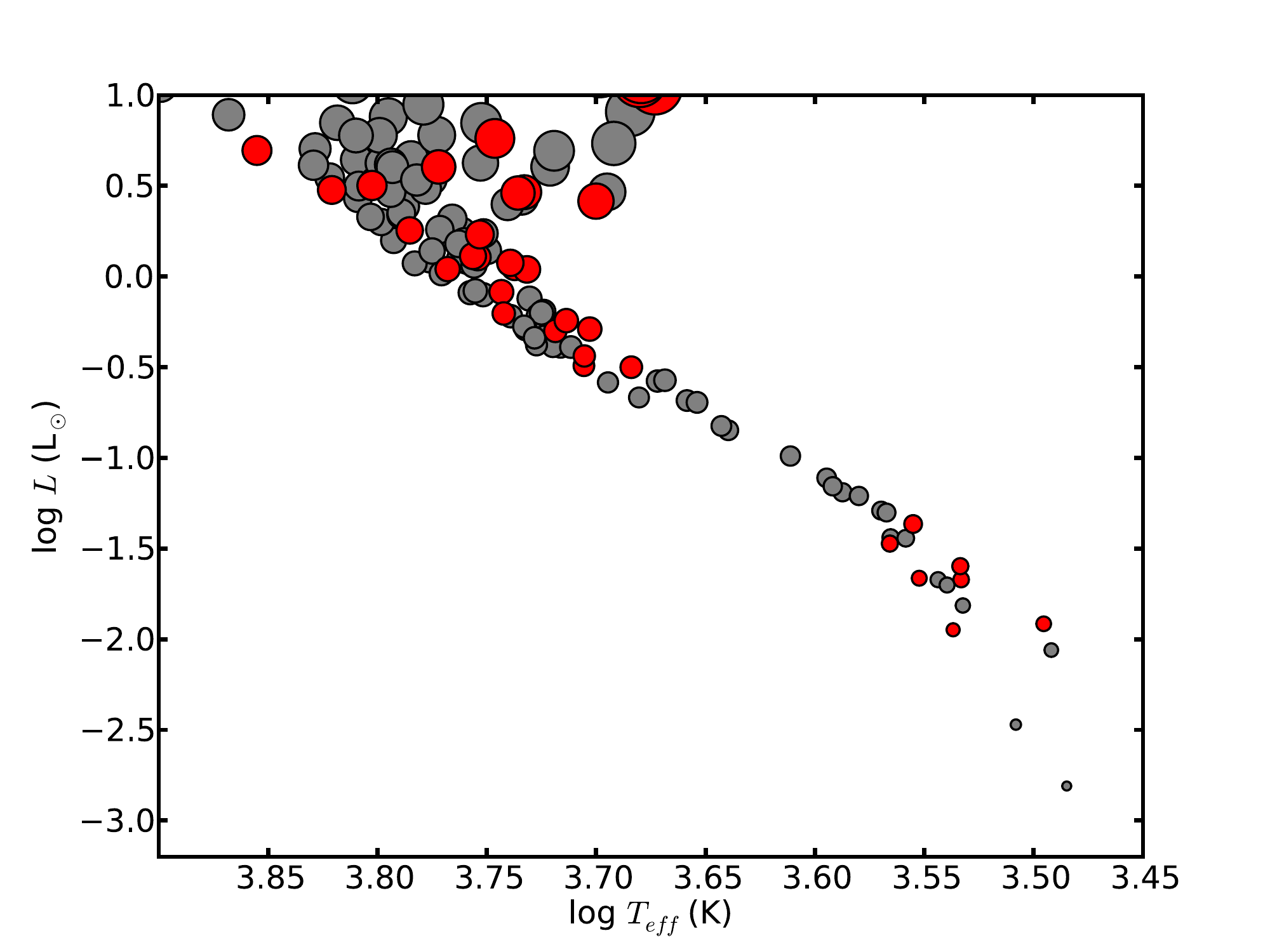}
\caption{Empirical H-R Diagram of the cool stars region for all stars with interferometrically determined stellar radii whose random uncertainties are smaller than 5\%. The diameter of each data point is representative of the logarithm of the corresponding stellar radius. Error bars in effective temperature and luminosity are smaller than the size of the data points. Exoplanet host stars are shown in red; stars that are not currently known to host any exoplanets are shown in grey. Stellar radii data are taken from \citet{bai08, bai12, bai13, big06, boy08, boy12, boy13, dif04, dif07, hen13, ker03, lig12, ric05, van99, van09, von11a, von11b, von12, von14, whi13}. \label{fig:HRD}}
\end{figure}

\begin{itemize}

\item Determination of empirical stellar parameters to help constrain stellar models on the low-mass end. The results confirm the well-documented discrepancy between theoretical and empirical stellar diameters for stars with effective temperatures lower than 5000K \citep[e.g.,][]{boy12, boy13}. 
\item Calibration of relations between stellar diameters and observables such as broad-band stellar colors \citep[e.g.,][]{boy14}.
\item Characterization of astrophysical parameters of exoplanets, as well as the system habitable zones, via the study of the host star \citep[e.g.,][]{von11a, von11b, von14}. 
\item For transiting planet systems, full characterization of stellar and planetary astrophysical parameters, including masses and bulk densities \citep[e.g.,][]{von12}.

\end{itemize}

Figure \ref{fig:HRD} shows the cool stars region of the empirical H-R Diagram based on interferometrically determined stellar radii from a number of different studies.

\section{Kepler's Cool Eclipsing Binaries: the Low Hanging Grapefruit}
Precise light curves from {\kepler} have motivated a comprehensive program to measure the empirical relationship between the masses, radii and effective temperatures of low-mass stars as a function of their 
absolute magnitudes and rotation periods. This program is designed to resolve 
the discrepancies regarding how rotation and stellar activity may affect the 
physical parameters of low-mass stars \citep[{\eg,}][]{Mullan2001,Chabrier2007,Kraus2011}, 
and will constitute a benchmark dataset for determining the physical parameters 
of exoplanet host stars for existing datasets such as {\kepler}, as well as 
upcoming space missions such as K2 \citep{Howell2014}, TESS \citep{Ricker2014}, 
and the James Webb Space Telescope.

This study is based on the high-precision lightcurves from the {\kepler} Space 
Mission \citep{Borucki2010,Koch2010,Jenkins2010} which provides over four 
years of nearly continuous photometric monitoring of approximately 4000 low-mass 
stars \citep{Dressing2013}. This sample of cool stars was searched for eclipsing 
binaries (EBs) using a custom implementation of a popular matched-filter algorithm 
\citep{Kovacs2002}, as well as Fourier transform techniques. The candidates from 
this search were inspected visually and cross matched against existing EB 
catalogs\footnote{\url{https://archive.stsci.edu/kepler/eclipsing_binaries.html}}\footnote{\url{http://keplerebs.villanova.edu}} 
to vet false positives, most commonly due to light leakage from other 
known EBs. The final list includes 24 EBs with periods longer than 1 day, 15 of which have 
periods longer than 3 days and extend out to 50 days. The rotation periods for the primary 
components (and in some cases, the secondary component) are measured from the Fourier transform of 
the {\kepler} light curves, else by use of the auto-correlation function 
\citep[{\eg},][]{McQuillan2013}. 

\begin{figure}[!ht]
\centering
\includegraphics[angle=0,width=4.8in]{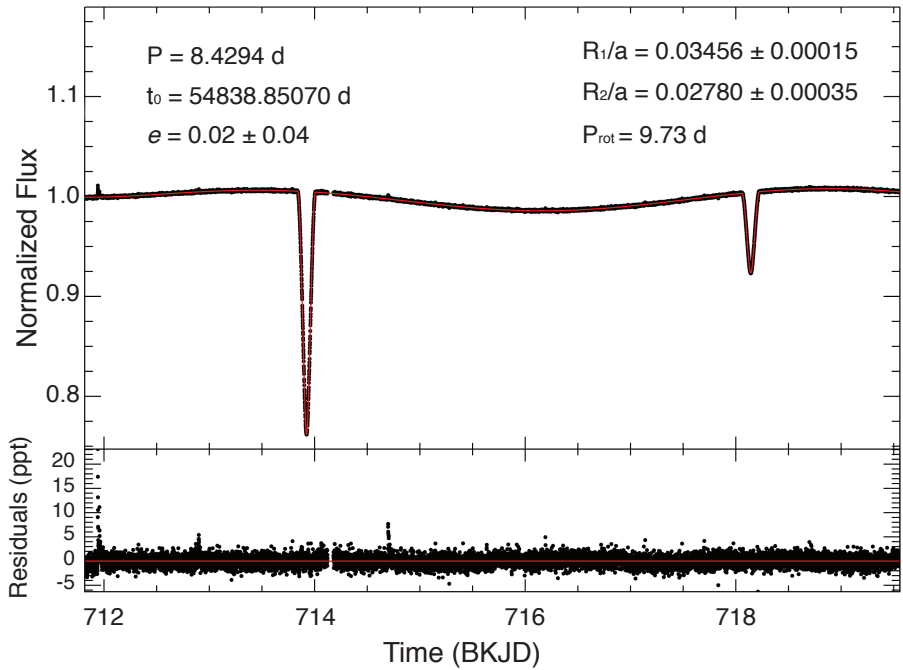}
\caption{A single eclipse cycle for of one of the brighter EBs in this sample. The {\kepler} short
cadence photometry (black dots) is overlaid with the best fit model (red curve), and the residuals 
are show in units of parts-per-thousand below. Many such fits (hundreds) can be combined to attain 
higher precision and accuracy on the derived stellar paramters.}
\label{fig}
\end{figure}

The eclipse light curves are modeled using the {\sc jktebop} code
\citep{Nelson1972,Popper1981,Southworth2004a} which, in its latest implementation,
allows for the simultaneous modeling of radial velocity measurements, accounts for the 
smoothing of the light curve due to finite integration times, and includes the capability to 
model the modulation of the light curve due to spots from the primary, secondary or 
third light components \citep{Southworth2013}. The high precision of the {\kepler} data 
allows the scaled radii of the M dwarf EBs ($R_1/a$ and $R_2/a$) to be determined to 
better than 1\% accuracy, and in many cases to better than 0.1\%. An example of one  
fit to a single primary/secondary eclipse pair for one of the brighter EBs in this sample is shown in Figure~\ref{fig}. Results show that for cases in which the 
rotation periods of the stars are not commensurate with the binary period that the effects 
of spot modulation on fitted parameters average out over the time baseline of the {\kepler} 
dataset. This allows for a much simpler approach to fitting where the lightcurves can be 
``detrended'' and normalized before fitting the entire dataset.

An approved observational program is currently underway to measure the 
radial velocities for the entire sample using the W. M. Keck Observatory with sufficient 
accuracy and phase coverage to determine the masses of both components to better than 3\% 
accuracy. Also ongoing are multi-band near-infrared and optical eclipse measurements 
with the Perkins $72^{\prime\prime}$ Telescope at Lowell Observatory and the MINERVA test 
facility\footnote{\url{http://www.astro.caltech.edu/minerva}} on the Caltech campus, 
respectively. From these observations, measurements of the effective temperatures of both 
stellar components can be made. A parallax program using the Discovery Channel 
Telescope at Lowell observatory will determine the absolute distances to the closest EBs in the 
sample, while the distances to the rest of the sample will have to wait for Gaia parallaxes.

\section{Constraining the early stages of binary star evolution with eclipsing binaries}

Detached, double-lined eclipsing binaries (EBs) are extremely valuable objects because their masses and radii can be determined in a model independent manner to a precision of a few percent or less \citep{Andersen91,Torres2010}. This makes them excellent, independent tests of models of stellar evolution. While there are many well-characterized systems on the main sequence (and hence many observational constraints), there are only 9 such systems on the pre-main sequence (PMS) below 1.5\,$M_{\odot}$. Constraints on the PMS are important because evolution is very rapid and the models are still sensitive to their initial conditions. The lack of low-mass, PMS EBs is well known and many efforts have tried to address their paucity \citep[e.g.][]{Aigrain07,Morales-Calderon12}. 

Detecting and characterizing a sample of low-mass EBs sharing the same age and composition, yet spanning a wide range of masses, was one of the key motivations for CoRoT to observe the $\sim$3 Myr old NGC\,2264 star forming region for 23 days in 2008. Thirty-seven EBs were identified among the possible cluster members, which were then targeted with an intensive program of optical ground-based spectroscopic follow-up to confirm their membership and determine their fundamental parameters. Masses and radii for both stars were derived by modeling the light curve and radial velocity data. Innovative Gaussian process regression techniques were utilized as part of this process in order to model both the out-of-eclipse variability in the light curves and the correlated noise in the cross-correlation functions when determining radial velocities (see \citealt{Gillen14} for further details).

\begin{figure}[h!]
  \centering  
  \includegraphics[width=4.8in]{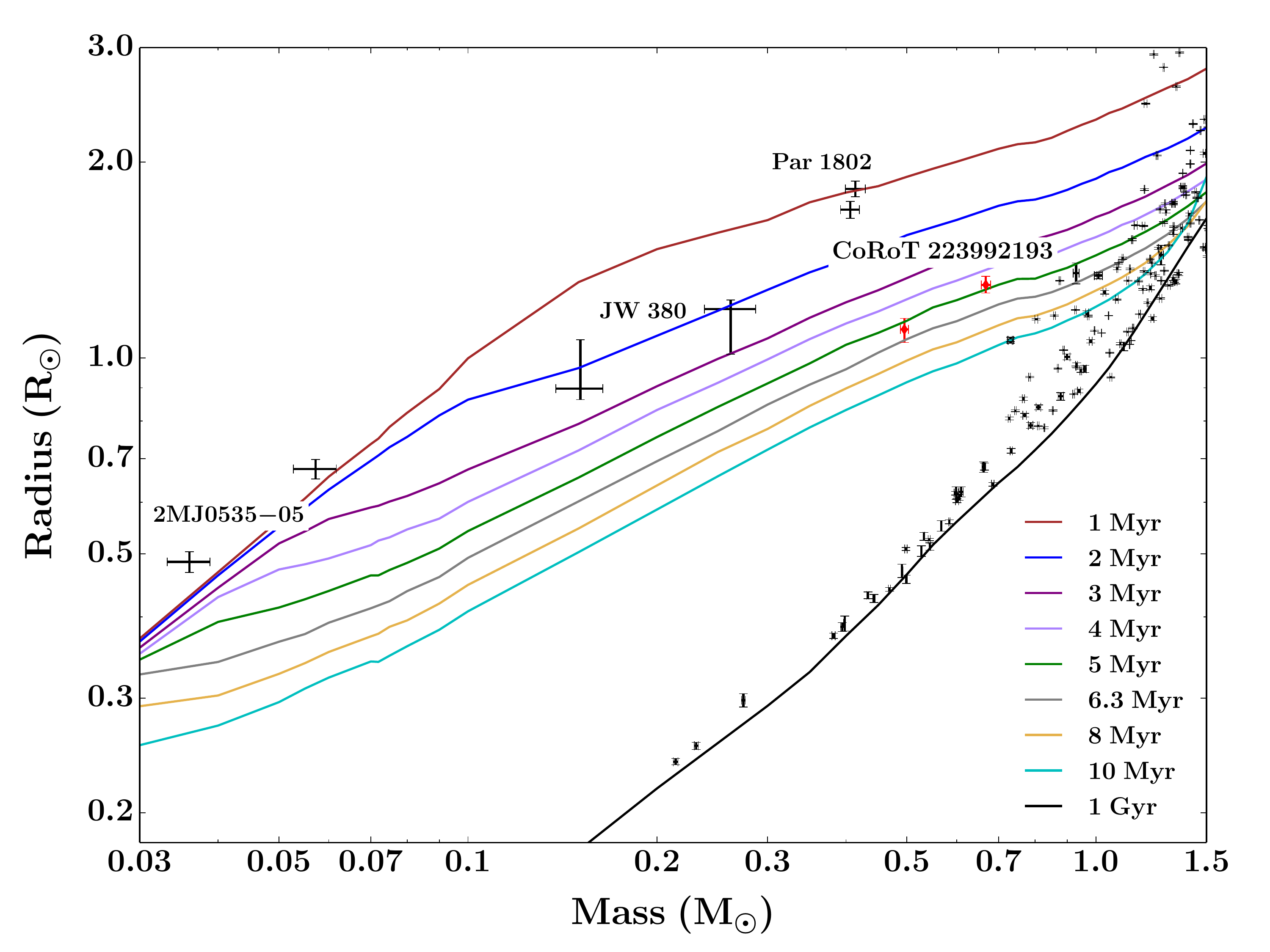}
   \caption[]{Mass-radius relation for low-mass EBs. The black points show measurements for stars in detached EBs\footnotemark, and the coloured lines show, from top to bottom, the theoretical isochrones of \citet{Baraffe98} for 1, 2, 3, 4, 5, 6.3, 8, 10\,Myr and 1\,Gyr (brown, blue, purple, lilac, green, grey, ochre, cyan and black, respectively. $Y=0.282$, $[{\rm M}/{\rm H}]=0$, mixing length $\alpha = 1.9$). The newly confirmed system, CoRoT 223992193, is plotted in red. For comparison, the three lowest mass systems known in the Orion Nebula Cluster are also labelled. Figure taken from \citet{Gillen14}.}
   \label{MR_EB_plot}
\end{figure}
\footnotetext{Data from John Southworth's catalogue, \tt{http://www.astro.keele.ac.uk/$\sim$jkt/debdata/debs.html}.}

This work has enabled the confirmation of one PMS cluster member. This system is shown in red in Figure \ref{MR_EB_plot}, which shows the mass-radius relation for low-mass, detached EBs. The system lies in a sparsely populated region of the diagram highlighting its importance in constraining models of PMS stellar evolution. Both components lie on roughly the same isochrone suggesting an apparent age of $\sim$5 Myr and the small error bars highlight the power of space photometry. This system, CoRoT\,223992193, is particularly interesting as it also shows evidence for a circumbinary disk \citep{Gillen14}. If confirmed, this would be the first such system discovered and could be an early analogue of the Kepler circumbinary planetary systems \citep[e.g.][]{Doyle11,Welsh12,Orosz12,Orosz12a}.

Seven other detached, double-lined EBs were identified, which are most likely field systems, and 19 single-lined EBs, four of which could be cluster members. The latter are low-to-extreme mass ratio systems where the primary star dominates the optical light and hence characterisation of the secondary is more favourable in the IR. Two of these four potential cluster members are currently being followed up with Gemini/GNIRS and one has been observed three times to date. A preliminary solution confirms its youth and hints at a large age discrepancy between the primary and secondary components. If this discrepancy persists with the inclusion of extra data and a more thorough analysis, it could be a very interesting system for star formation theories in cluster environments.

Looking to the future, the Kepler mission is continuing in 2-wheel mode (K2; \citealt{Howell2014}) and, over the next two years, will observe a number of star forming regions and young clusters ranging from $\sim$1 to several hundred Myr. These observations, along with ground-based spectroscopic follow-up, promise to help populate the PMS with well-characterised EBs that can act as powerful observational constraints on models of stellar evolution.

\section{Temperatures and radii of low mass dwarfs inferred from NIR spectra}
While direct measurements of stellar properties are only within reach for a small sample of nearby, bright dwarfs, touchstone stars are an invaluable sample from which to develop empirical relationships for fundamental stellar properties, which may then be applied to fainter and more distant stars. Results from CHARA have greatly increased the number of low mass dwarfs with interferometrically measured radii \citep[e.g.][]{boy12, von14}. \citet{Newton2014a} used these touchstone stars to calibrate relationships for the radii, effective temperatures ($\teff$), and log bolometric luminosities of cool dwarfs with radii between $0.2$ and $0.8\rsun$. These calibrations use the EWs of strong Mg and Al features in $R\sim2000$ H-band spectra. The standard deviations in the residuals of the best fits are $0.03\rsun$, $70$K, and $0.05$~dex, respectively.

\begin{figure}
\includegraphics[width=4.8in, trim=100 350 100 100]{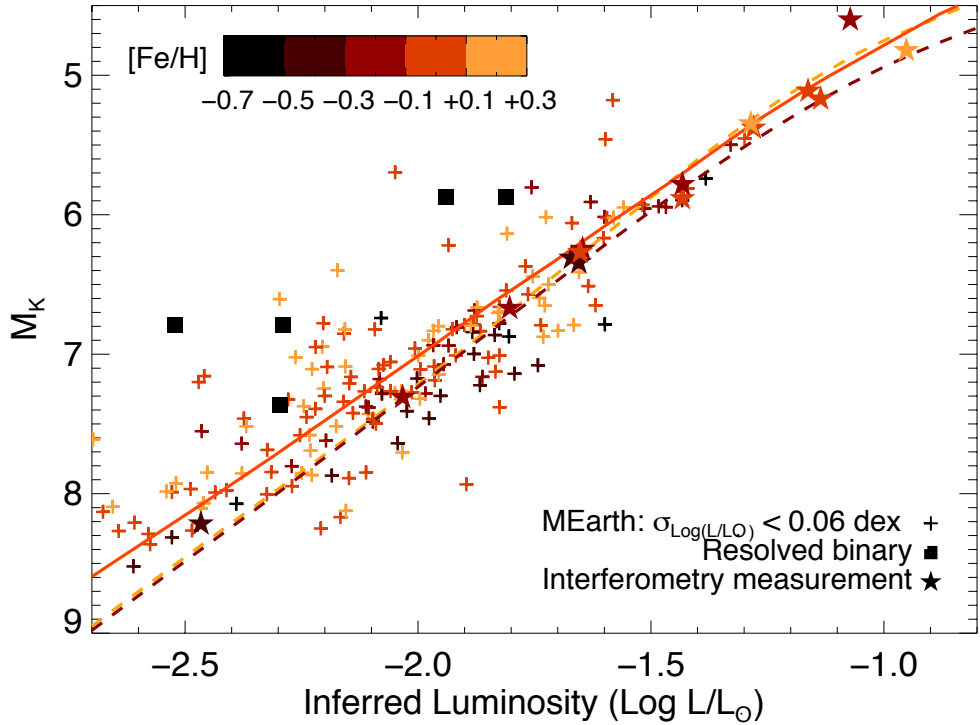}
\caption{Inferred luminosity versus absolute K magnitude for stars with errors on $\mk<0.1$ mag. The BCAH \citep[][red solid line]{Baraffe98} and Dartmouth \citep{Dotter2008a} 5 Gyr isochrones are over plotted. Two metallicities are shown for the Dartmouth isochrones: $\mathrm{[M/H]}=+0.3$ (orange dashed line) and $\mathrm{[M/H]}=-0.5$ (dark red dashed line) and the measured properties of the interferometry stars (filled stars). The $\feh$, estimated using the relation from \citet{Newton2014}, is indicated by its color. Binaries that are unresolved in 2MASS but resolved in spectral observations are also indicated (filled black squares). \citet{Newton2014} identify stars with brighter $\mk$ than expected given the inferred luminosity as candidate multiples.
\label{Fig:newton}}
\end{figure}

\citet{Newton2014a} applied their calibrations to the sample of M dwarf spectra obtained by \citet{Newton2014}, 90\% of which have parallaxes either collected from the literature or measured from the MEarth transiting planet survey by \citet{Dittmann2014}. They found that the radii and luminosities they infer vary with absolute K magnitude as expected from stellar models, as shown in Figure \ref{Fig:newton}. They find that visual binaries, for which independent spectra were acquired of each component but which are unresolved in 2MASS, tend to have brighter K magnitudes that one would expect from their inferred luminosities. They use this over-luminosity to identify candidate multiples; half of the candidates identified are known binaries that were unresolved in the spectral observations.

Using H-band spectra of Kepler Objects of Interest from \citet{Muirhead2014}, \citet{Newton2014a} inferred the properties of cool stars hosting candidate planets. For stars with effective temperatures below 3900~K, the effective temperatures they infer agree well with those inferred in previous works. The stellar radii they infer from H-band agree with those from \citet{Mann2013a}, who base their work on a similar set of stars with interferometric measurements. However, they are larger than those inferred by \citet{Muirhead2014} and \citet{Dressing2013}, by $0.06\rsun$ and $0.09\rsun$, respectively who based their estimates by comparison with the PHOENIX \citep{Allard2011} and DSEP \citep{Dotter2008a} models, respectively. This has a direct effect on the properties of the inferred planets: the planetary radii one infers using the H-band stellar radii are larger by 14\% than those one would infer using the synthesized stellar parameter catalog from \citet{Huber2014}. This highlights the importance of tying inferences of stellar properties to direct observations of touchstone stars.

\section{Dating middle-aged stars with Ruprecht 147}

We have known for over 40 years that stellar rotation, magnetic activity and lithium abundances decay over time \citep{Skumanich1972}, 
making these properties potentially powerful age diagnostics 
for Sun-like stars, which otherwise change very little during their 
main sequence lives. It is widely assumed that our ability to empirically calibrate these age relationships is hindered by a lack of old, nearby open clusters \citep{Soderblom2010}.
Based on this understanding, astronomers have sought out 
additional age benchmarks, including wide binaries where
Sun-like dwarfs are paired with $>$MSTO stars \citep{Soderblom1991} 
or white dwarfs \citep{Chaname2012},
nearby Solar twins,
or asteroseismic targets \citep{Garcia2014}
\footnote{\citet{Mamajek2014} provides a detailed summary 
of rotation evolution for Solar mass stars at 
http://figshare.com/articles/The\_Sun\_Rotates\_Normally\_for\_its\_Age/1051826}.
These efforts are providing useful data points,  
but it is important to consider how definitions of proximity are relative, 
and change over time as observational techniques change 
and instrumentation improves. 

\textit{Kepler} has brought distant open clusters into view, 
where rotation periods have been measured for the 1 Gyr NGC 6811 \citep{Meibom2011}, 
and the 2.5 Gyr NGC 6819 (Meibom, in progress
\footnote{Initial results presented at the 400 Years of Stellar Rotation Workshop, 
see \citet{StellarRotation} for a conference summary}). 
The \textit{K2} mission has been scheduled to observe 
many additional open clusters \citep{Howell2014},
including the 3 Gyr Ruprecht 147 (hopefully)
and 4 Gyr M67 benchmark clusters.
Combining the \textit{Kepler / K2} open clusters with a feasible ground based 
gyro survey of the 1.6 Gyr NGC 752 (450 pc) could provide large samples of stars 
with masses spanning the main sequence, for clusters older than the Hyades at 
1 Gyr (NGC 6811), 1.6 Gyr (NGC 752), 2.5 Gyr (NGC 6819), 
3 Gyr (Rup 147) and 4 Gyr (M67). 
These rotation periods may well demonstrate the level of star-to-star 
scatter inherent to angular momentum evolution, 
along with the empirically-calibrated functional relationship 
for gyrochronology\index{gyrochronology} in Sun-like stars.

Ruprecht 147\index{Ruprecht 147} (NGC 6774) is the oldest nearby star cluster, 
with an age of 3 Gyr at 300 pc, 
which allows it to serve as a sorely needed intermediate-aged benchmark 
\citep{Curtis2013}.
Rup 147's proximity has allowed us to 
measure coronal X-ray luminosities from Sun-like stars \citep{Saar2014};
spectroscopically identify and characterize the white dwarf and red dwarf populations
(work in progress);
and obtain high-resolution spectra for nearly the entire membership, 
enabling studies of chromospheric activity (Curtis et al. 2014b, in prep), 
lithium depletion (Ram{\'{\i}}rez et al. 2014, in prep), 
and detailed stellar characterization (Curtis et al. 2014a, in prep). 
Ruprecht 147 is located in Sagittarius, just beyond the edge of the proposed field for \textit{K2} Campaign 7, but the pointing was adjusted in response to community advocacy to accommodate the cluster. Observations are scheduled for late 2015\footnote{http://keplerscience.arc.nasa.gov/K2/Fields.shtml\#7}. The cluster's proximity will allow asteroseismic analysis of the 
red giant, and possibly upper main sequence turnoff, stars
(Victor Silva Aguirre, private communication), 
rotation periods for the Sun-like stars, 
and $\sim$3 \rearth\ exoplanet detection around the M0 dwarfs \citep{Howell2014}.

\begin{figure}\begin{center}
\plotone{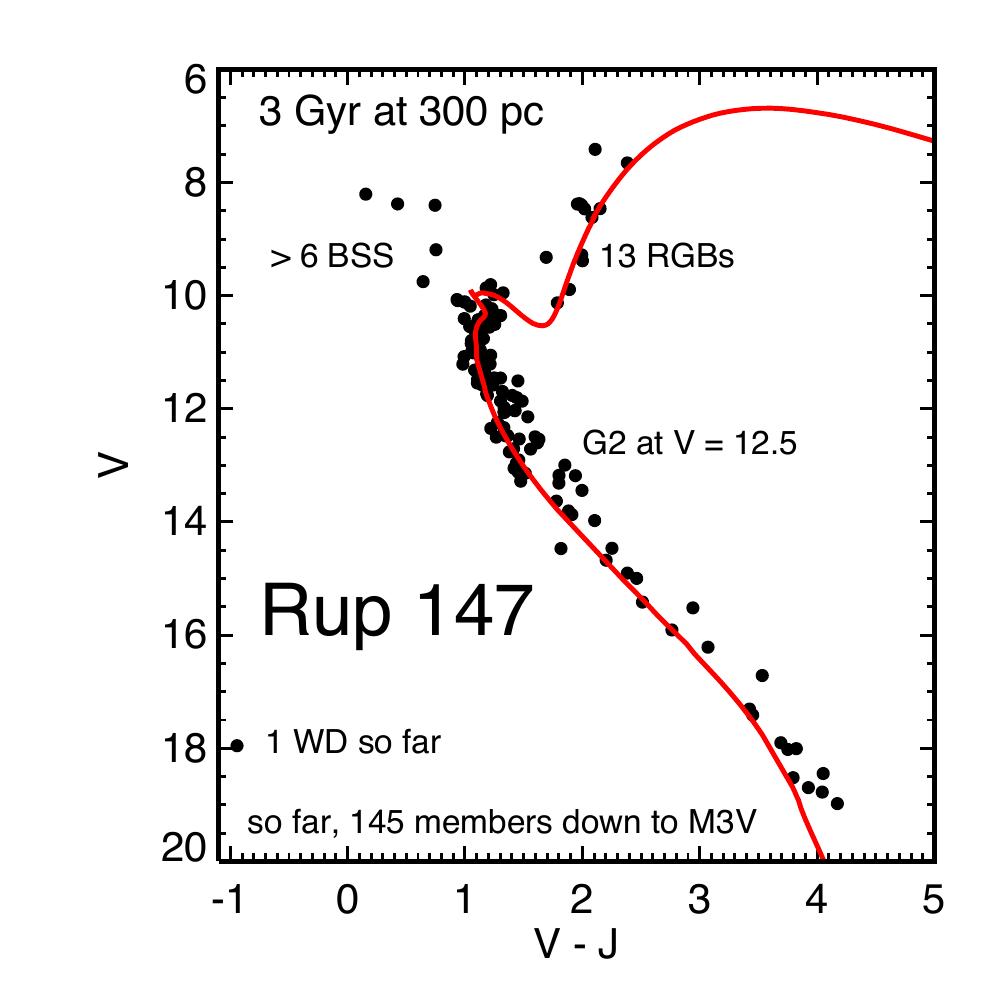}
	\caption{Color-magnitude diagram for Ruprecht 147, along 
	with 3 Gyr Dartmouth isochrone with [Fe/H] = +0.1, $A_V = 0.25$ 
	at 300 pc.
	\label{f:cmd}}
\end{center}\end{figure}

There is still a gap in the magnetic activity data 
between the age of the Hyades, and M67 and the Sun. 
While \citet{Mamajek2008} demonstrated that magnetic activity 
declines continuously with time, 
this picture has been called into question by the recent 
high-quality VLT data for distant, 1-2 Gyr clusters 
presented by \citet{Pace2004} and \citet{Pace2009}. 
While their samples are small, their data suggest 
an abrupt decline in chromospheric emission
after 1 Gyr, with little to no subsequent evolution. Ruprecht 147's Solar analogs are all 
more active than the active Sun, as expected from \citet{Mamajek2008}, 
but possibly still consistent with an abrupt decline around 1-2 Gyr. There is an ongoing chromospheric survey of NGC 752 that will yield a large sample of intermediate-aged stars, which will help differentiate between these activity-evolution scenarios. 
Further progress on this question will require larger samples of well-characterized stars spanning 1--3 Gyr,  that can complement the cluster data. The Solar analogs with rotation periods between 11-20 days observed by \textit{Kepler / K2} are natural targets.

We have not yet exhausted the observable middle-aged clusters; in fact, we still await most of the potential observations and rotation periods. 

\section{Expanded SEDs and Bolometric Luminosities as Direct Measures of Substellar Touchstones }

Unable to sustain nuclear fusion in their cores due to insufficient mass, all brown dwarfs are about the radius of Jupiter, emit primarily in the infrared, and are degenerate across effective temperature, mass, and age. These observational challenges make stellar direct measurement methods such as interferometry and asteroseismology unfeasible and dynamical mass measurements very difficult for a large, diverse sample. However, spectral energy distributions (SED) with optical through mid-infrared data allow detailed spectroscopic study of these objects over a broad wavelength baseline. With the addition of parallaxes, precise bolometric luminosities ($L_{bol}$) become a robust direct measurement by which one can investigate the effects of different fundamental parameters on the global characteristics of brown dwarfs.

Near-complete SEDs for 175 stars are constructed, covering the entire sequence of late-M, L and T dwarfs, using optical and near-infrared (NIR) spectra from the BDNYC Data Archive combined with mid-infrared (MIR) data from the Wide-field Infrared Survey Explorer (WISE; \cite{Wright_2008}) and the Spitzer Space Telescope (Fazio et al. 2004; Houck et al. 2004). Objects with optical signatures of low surface gravity ($\beta$ or $\gamma$; \cite{Kirkpatrick_2005}, \cite{Cruz_2009}) or membership in nearby young (10-150 Myr) moving groups (NYMGs; Faherty et al. in prep) are identified as 24 percent of the sample to investigate the effects of temperature, gravity, and dust/clouds on spectral morphology. $L_{bol}$ is calculated for this sample by integration of the SEDs, flux calibrated using parallax measurements from the Brown Dwarf Kinematics Project (\cite{Faherty_2012}) and the literature (\cite{Dupuy_2012}, \cite{Tinney_2003}, \cite{Vrba_2004}) or published kinematic distances (\cite{Cruz_2003}; \cite{Faherty_2009}; \cite{Schmidt_2007}, \cite{Schmidt_2010}; \cite{Reid_2006}; \cite{Delorme_2012}; \cite{Naud_2014}).

\begin{figure}[h!]
\begin{center}
\includegraphics[width=4.8in]{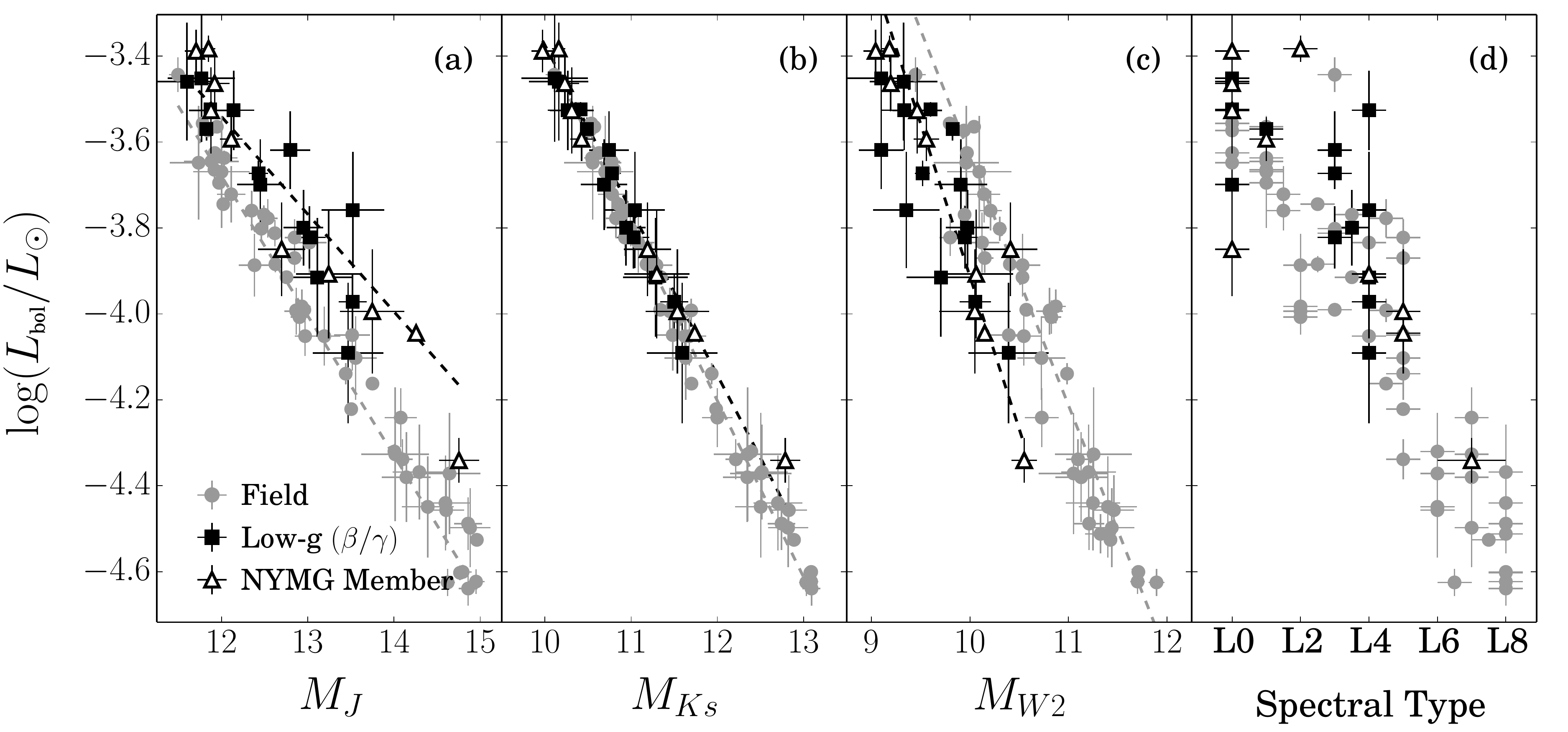}
\caption{Bolometric luminosities of field age (light filled circles), low surface gravity (dark filled squares), and members of nearby young moving groups (unfilled triangles) are plotted against absolute J, Ks and W2 magnitudes and spectral type for a selection of 54 L dwarfs. First order polynomials fits to the field objects (light dashed line) and the young and low gravity objects (dark dashed line) show distinct tracks in $M_J$ and $M_{W2}$ and overlapping tracks in $M_{Ks}$.}
\label{fig:gillen}
\end{center}
\end{figure}

Figure~\ref{fig:gillen}(a,b,c) shows $L_{bol}$ versus selected absolute magnitudes $M_J$, $M_{Ks}$ and $M_{W2}$ for 54 field age, 14 low-g, and 10 NYMG member L dwarfs of the sample. The flux of low surface gravity (low-g) L dwarfs appears to be redistributed from the NIR into the MIR, primarily from J to W2, as compared to field age Ls of the same luminosity (\cite{Faherty_2012}, \cite{Faherty_2013}; \cite{Liu_2013}; \cite{M_R_Zapatero_Osorio_2014}; Gizis et al., submitted). Indeed low-g Ls appear 0.5-1 magnitudes dimmer in $M_J$ and 0.3-0.6 magnitudes brighter in $M_{W2}$ (Filippazzo et al., in prep). This is probably due to absorption and scattering of light to longer wavelengths by diffuse, unsettled dust in the atmospheres of young objects. Additionally, $M_{Ks}$ magnitudes appear to be largely unaffected by surface gravity making it an ideal band from which to determine age-independent bolometric corrections for L dwarfs. 

The plot of $L_{bol}$ versus spectral type (Figure~\ref{fig:gillen}d) shows most low-g, young, and field objects all lie along the same sequence. Qualitatively, low-g L dwarfs have larger radii than their field age counterparts of the same $L_{bol}$ so they must have cooler photospheres according to the Stefan-Boltzmann Law. Bolometric luminosities are one of the few direct measurements one can make for brown dwarfs for identification of substellar touchstones, however, effective temperatures ($T_{eff}$) can also be tightly constrained using evolutionary models while minimizing assumptions about the sources (See Filippazzo et al., in prep). 

Preliminary results suggest that confirmed young objects are 100-400 K cooler than field age L dwarfs of the same spectral type. While the uncertainties on the radii of low-g (but not necessarily young) objects are large, they still fall below the track of "normal" Ls on the $T_{eff}$ versus spectral type plot (Filippazzo et al. in prep). Consequently, surface gravity must be taken into account when using spectral type as a proxy for $T_{eff}$. The reliance on an age insensitive temperature-spectral type relationship (\cite{Golimowski_2004}, \cite{Stephens_2009}) might explain why some young objects appear underluminous as compared to older L dwarfs of the same temperature.

Though the complex nature of their cool atmospheres obscures the fundamental parameters of brown dwarfs, the picture of substellar touchstones becomes clearer as larger age-calibrated samples are assembled and detection methods improve. 

\section{Photometric Calibration for Determining M-Dwarf Metallicity}
M dwarfs may be small, cool and faint but they are the most numerous stars in the Galaxy. Their Main Sequence lifetimes are much longer than the current age of the Universe and they can, therefore, be used as excellent tracers of Galactic structure and population as well as Galactic chemical, kinematical and dynamical evolution.  However, a full understanding of stellar and Galactic astronomy requires accurate knowledge of the fundamental parameters of these dwarfs, such as metallicity.  Although there has recently been significant progress in deriving empirical spectroscopic methods through moderate-resolution spectra, photometric analyses provide more efficient techniques to determine the metallicity of large numbers of M dwarfs for Galactic studies. Instead, a photometric method to measure metallicities could readily applied to large numbers of stars without the need for parallaxes, moderate-to-high resolution spectra or time intensive calculations. 

\begin{figure}
\includegraphics[angle=0,width=4.8in]{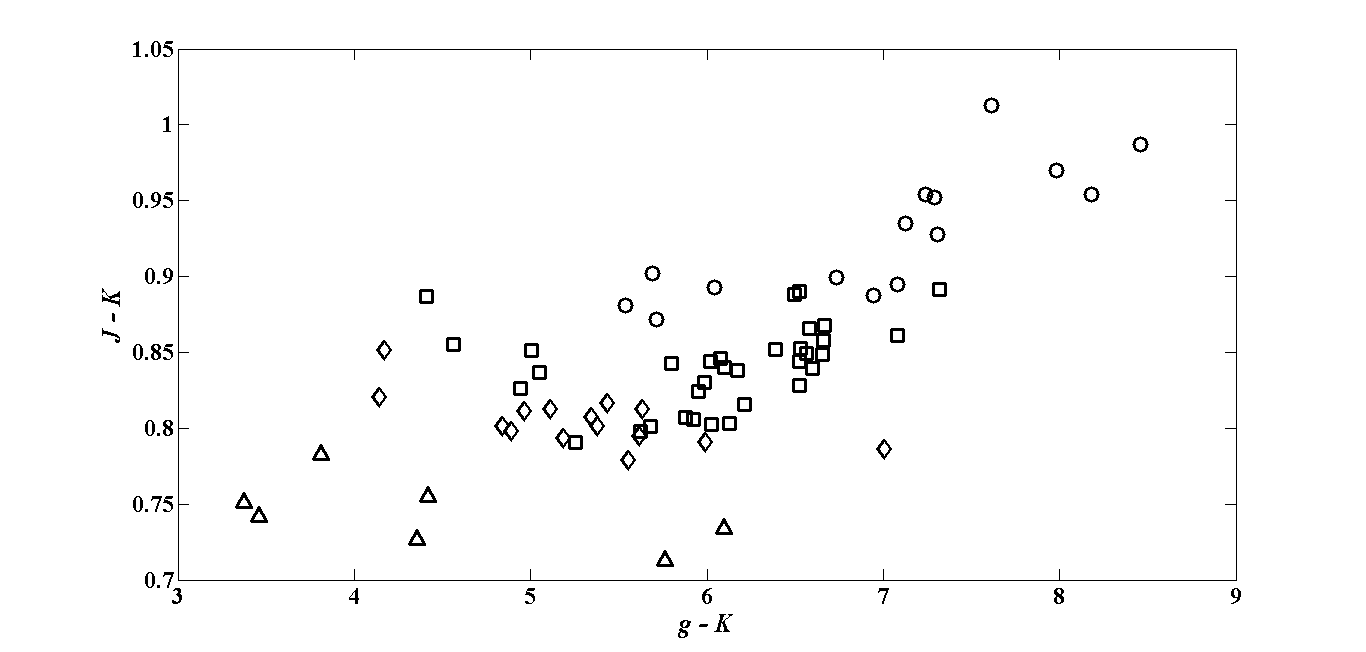}
\caption{The ({\it{J-K}})-({\it{g-K}}) color-color diagram of the 71 M and late-type K dwarfs in the calibration sample.The symbols indicate the metallicity range of stars: [Fe/H] $\geq$ 0.15 dex as circles, -0.1  $\leq$ [Fe/H]  $<$ 0.15 as squares,   -0.4  $\leq$ [Fe/H]  $<$ -0.1 as diamonds, and [Fe/H] $<$ 0.4 as triangles}\label{fig:jhk}
\end{figure}

To calibrate metallicity, 21 M dwarfs were identified that have trustworthy SDSS {\it{g}} and 2MASS {\it {JHK}} photometry companions to an FGK star or early-type M dwarf of known metallicity. It was found that among all possible colors, the {\it{J-K}} color of these stars has the best metallicity sensitivity and the ({\it{J-K}})-({\it{g-K}}) color-color diagram can effectively distinguish metal-poor dwarfs from metal-rich ones. Fifty other M dwarfs were added with reliable {\it{gJK}} photometry and spectroscopically determined metallicities through moderate-resolution spectra. These stars were selected based on their metallicities and locations on the ({\it{J-K}})-({\it{g-K}}) diagram with respect to the 21 well-calibrated stars. The calibration sample then includes a total of 71 stars whose color-color diagrams are shown in Figure~\ref{fig:jhk}. By fitting the ({\it{J-K}})-({\it{g-K}}) plane using a low-order polynomial, relation between metallicity and colors was derived, yielding an RMSE=0.078 dex. This calibration can be applied to dwarf stars of spectral types between K6 and M6.5, with -0.73 $\leq$ [Fe/H] $\leq$ +0.3 dex, 3.37 $\leq$ {\it{g-K}} $\leq$ 8.46 and 0.71 $\leq$ {\it{J-K}} $\leq$ 1.01.

A sample of $\sim$300,000 M and late-type K dwarfs were identified from the matched catalog between SDSS DR9 and 2MASS surveys with Galactic latitudes $b$ $\geq$ 50$^\circ$ and Galactic longitudes 170$^\circ$ $\leq$ $l$ $\leq$ 210$^\circ$. To select this sample, the typical color ranges for M and late-type K dwarfs were used. Some other color cuts were also applied; those required for 1) the calibration sample, 2) removing giant stars, and 3) applying a recent photometric parallax technique to estimate the distance of stars. The metallicity and Galactic height of dwarfs in this large sample were determined using their photometry. Results show a shift in metallicity distribution toward low metallicities as the Galactic height increases. This can be attributed to the age-metallicity-Galactic height relation. Studies have shown that stars which formed at earlier times of the Galaxy's history generally have lower metallicity and are farther from the Galactic plane. Several models of Galactic chemical evolution were also tested in this study. It turns out that the Simple Closed Box Model overpredicts the number of metal-poor M dwarfs, which indicates the existence of the M-dwarf problem. It has been shown that models with declining inflow rates are elegant solutions to the local G-dwarf problem, as gas accretion to galaxies is a common phenomenon in the Universe. Several models of this kind were investigated relative to the distribution and found that all of these models can mitigate to some extent the M-dwarf problem.

\section{Confronting Predictions of Stellar Evolution Theory with Touchstone Stars}

We are in an age of precision stellar astrophysics. Fundamental stellar properties---mass, radius, T$_{\rm eff}$, luminosity---can be determined with precisions often better than 2\% on certain stars. These ``touchstone'' stars provide a rich sample with which to confront predictions of stellar evolution theory and thereby test stellar model physics. Observations of single stars and eclipsing binaries (EBs) have revealed
significant discrepancies between predictions from stellar models and the
reality portrayed by this sample of touchstone stars \citep[e.g.,][]{Torres2010,
boy12}. Real stars appear larger and cooler than theory predicts, a 
result that has been shown to be robust against variations in metallicity and age 
\citep{FC12,Spada2013}. There is no \emph{a priori} reason that stars in EBs and 
single stars should exhibit the same deviations from theoretical predictions, 
yet they do, suggesting general shortcomings in stellar models physics. 
Though often viewed in a negative light, model shortcomings highlighted by
touchstone stars present an exciting opportunity to explore new stellar atmospheric physics.

One of the more popular hypotheses, at the moment, is the presence of strong
magnetic fields \citep[e.g.,][]{Mullan2001} and/or magnetic activity
\citep[i.e., spots;][]{Chabrier2007}. The interaction between magnetic fields
and thermal convection (magneto-convection) is such that magnetic fields 
act to stabilize a fluid against convective instability, forcing radiation
to carry a greater amount of outgoing flux. To preserve total flux at its surface, the star must inflate (and consequently become colder). Stellar models including simplified magneto-convection prescriptions can reconcile models with observations of touchstone stars
in EBs \citep[e.g.,][]{MM11,MM14,FC13,FC14} and, recently,
in the population of single stars \citep{Malo2014}. Quantitative predictions
of the required surface magnetic field strengths are often consistent with
expectations based on empirical data \citep{Reiners2012a}, implying that magnetic
models may be capturing relevant physical processes despite simplified 
prescriptions. There remains discussion, however, about the validity of 
magnetic models in the fully convective regime. Interior magnetic field
strengths in excess of 1~megaGauss are typically required \citep{Chabrier2007,MM11,MM14,
FC14}, which may be difficult to produce in a fully convective dynamo and 
which threaten to quickly become unstable if they do develop. 

Star spots can be invoked to relieve the burden placed on models of magneto-convection, particularly in the fully convective regime
\citep{MM11}. \citet{Morales2010} demonstrated that spots can bias radius measurements
toward larger radii by as much as 6\%, if spots are located preferentially 
near the poles and cover a significant fraction (upward of 60\%) of the 
stellar surface. Slight radius inflation from suppression of flux leaving
the stellar surface is also expected, with the two effects conspiring to
explain observations of radius anomalies for fully convective EBs. Presently,
there is no strong empirical evidence of polar spots on rapidly rotating low 
mass stars or stars below the fully convective boundary, but relatively few
stars have been studied in this capacity \citep{FC14}. Further observational
evidence for or against the magnetic hypothesis is still needed. In the meantime,
studies are underway to explore other potential explanations. 
 
\begin{figure}
	\centering
	\includegraphics[width = 4.5in]{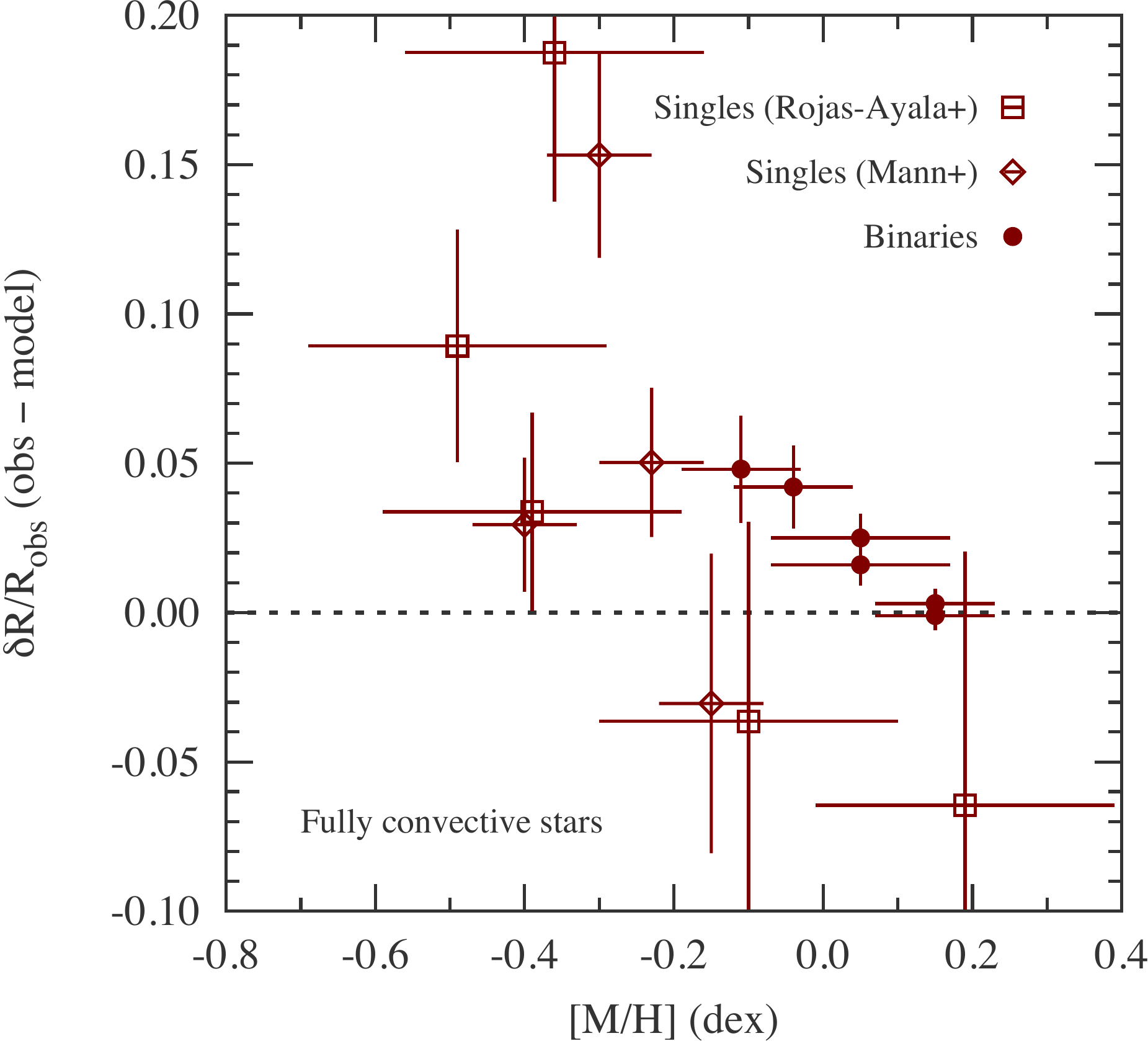}
	\caption{Relative error between radii of fully convective touchstone stars and 
		radii predicted by stellar evolution models. Stars in EBs are plotted as solid
		circles, while single stars with radii measured using interferometry are 
		shown with open symbols, the shape depending on the empirical calibration 
		used to determine the stellar metallicity: \citet[open squares]{RojasAyala2012}
		or \citet[open diamonds]{Mann2013b}.
		\label{fig:feh_rad}}
\end{figure}

If we instead look only at fully convective stars in EBs, a pattern emerges with 
lower metallicity stars displaying larger deviations from theoretical predictions 
(see Figure~\ref{fig:feh_rad}). With only six fully convective
stars in EBs that possess reliable metallicity estimates, it is difficult to draw a firm
conclusion that metallicity and radius deviations are anti-correlated, but it
is tantalizing nonetheless. Isolating the fully convective stars in the interferometric
sample makes this hypothesis less compelling, but the lack of mass estimates adds
uncertainty to unambiguously identifying stars below the fully convective boundary. 
However, \citet{boy13} have indicated that a trend with metallicity might
exist in the full interferometric sample (see their Figure~2). However, the sample shows considerable scatter and is sparsely populated at the low metallicity end. No compelling physical explanation has been offered for an anti-correlation between radius deviations and metallicity.

Already, touchstone stars have identified the possible need for new model physics
in the form of magnetic fields and/or activity, as well as the possible need for 
improvements in current micro-physics. We can only expect a more robust identification 
of model shortcomings as the sample of touchstone stars increases, leading to more 
directed refinements of current micro-physics and additions of new physics. With advancements 
in determining metallicities of low mass stars \citep[e.g.,][]{RojasAyala2012,Terrien2012a,
Mann2013a}, fully convective stars may provide one of the more interesting subsets
of touchstone stars. They are relatively insensitive to current model parameters 
and may therefore more easily reveal areas deserving of concerted attention. One
example is using remaining discrepancies between models and fully convective touchstone
stars to investigate helium abundance as a function of metallicity in the local galactic
neighborhood \citep[see, e.g.,][]{Ribas2000}.

\acknowledgments{
We thank the CS18 science organizing committee for selecting this splinter session to be part of the Cool Stars 18 conference and the CS18 local organizing committee for their logistical and technical support with the splinter session. 
}

\end{document}